\documentstyle[12pt]{article}
\def\be{\begin{equation}}

\def\ee{\end{equation}}
\def\bea{\begin{eqnarray}}
\def\eea{\end{eqnarray}}
\def\nn{\nonumber \\}

\def\be{\begin{equation}}

\def\ee{\end{equation}}
\def\bea{\begin{eqnarray}}
\def\eea{\end{eqnarray}}
\def\nn{\nonumber \\}

\begin{document}
\titlepage
\begin{flushright}
CERN-TH.7552/95 \\
OUTP 94-38P      \\
RAL-95-002      \\
January 1995
\end{flushright}
\begin{center}
\vspace*{1.0cm}
{\Large{\bf An analytic solution of the BFKL equation with momentum
cutoffs }}
\end{center}
\vspace*{.75cm}
\begin{center}
M.F.McDermott$^1$   \\
J.R.Forshaw$^2$  \\
and \\
G.G.Ross$^1$ \\
$^1$ Department of Theoretical Physics, University of Oxford,  \\
1-4 Keble Road, Oxford, OX1 3NP, United Kingdom. \\
$^2$ Rutherford Appleton Laboratory,\\ Chilton, Didcot, Oxon, OX11 0QX,
United Kingdom.\\
\end{center}
\vspace*{1.2cm}
\begin{abstract}
We outline a general method for obtaining the solution to the
($t=0$) BFKL equation in the presence of transverse momentum cutoffs.
A lower cutoff allows one to avoid integration over nonperturbative
momenta and an upper one is needed from energy-momentum conservation.
Our method allows for the inclusion of an arbitrary number of poles in the
kernel and is applicable to any input distribution. Taking Mellin transforms,
we discuss the effect of introducing cutoffs by considering  the
singularity structure in the transform space. We present an improved
calculation of the small-$x$ slope of the gluon density.
\end{abstract}
\newpage

\section{Introduction}
Recent results from the HERA electron-proton collider on the small-$x$ rise
of the nucleon structure function $F_2(x,Q^2)$ \cite{hera,wolf} have
generated considerable theoretical interest in small-$x$ physics.
 Theoretical opinion on the cause of this increase is somewhat divided
\cite{bf,bfkl,akms}.
One explanation is that HERA has become sensitive to a new regime
of  perturbative QCD, i.e. the `small-$x$ limit' defined by
\begin{eqnarray}
\ln(\frac{1}{x})    &  \gg    &  1  \nn
\alpha_s\ln(\frac{1}{x})  & \sim  & 1.
\end{eqnarray}
Generally, $x \sim Q^2/s$ where $Q^2 \gg \Lambda_{QCD}^2$ is some hard
 scale and $s$ is the process centre-of-mass energy.
 In the case of deep inelastic
scattering, $x$ is the Bjorken-$x$ and $Q^2$ is the photon virtuality.

Within this regime an equation has been derived, the  BFKL equation
\cite{bfkl}, which sums those terms in the perturbative expansion which
have equal powers of $\alpha_s$ and $\ln(1/x)$. In this limit, the
strong coupling is not renormalised and remains fixed.  The BFKL equation
(after angular averaging) can be expressed as an integral equation
for the  evolution of the unintegrated gluon distribution function
$f(x,k^2_{\perp})$, for momentum transfer squared $t=0$ it is given by

\begin{equation}
\frac{ \partial f( x,k^2_{\perp} ) }{\partial \ln 1/x} =
( \frac{ 3\alpha_s k^2_{\perp} }{ \pi } ) \int_{0}^{\infty} {\displaystyle
\frac{ dk^{\prime2}_{\perp} }{  k^{\prime2}_{\perp}  }  }
\left\{ {\displaystyle \frac{ ( f( x,k^{\prime2}_{\perp} )
- f( x,k^2_{\perp} )  ) }{ \mid k^{\prime2}_{\perp} - k^2_{\perp} \mid}
+ \frac{ f( x,k^2_{\perp} ) }{ \sqrt{ k^4_{\perp} +
4k^{\prime4}_{\perp} } } } \right\}.
\label{bfkl}
\end{equation}
Given an input distribution, $f(x_0,k_{\perp}^2)$ (which we take to be
independent of $x$), eq.(\ref{bfkl}) produces the distribution
at lower $x$, $f(x,k_{\perp}^2)$.

Although the equation embodies the complete leading $\alpha_s \ln x$
summation it is clear to see that there are potential difficulties
associated with the infra-red and ultra-violet regions of the
(semi-infinite) integration over $k_{\perp}^{2 \prime}$ in
eq.(\ref{bfkl}). For small values of $k_{\perp}^{2\prime}$,
we expect the corrections to the leading log resummation to be
important, e.g. the introduction (by hand) of the QCD running
coupling, evaluated at the scale
$k_{\perp}^2$, produces a logarithmic infra-red divergence in
eq.(\ref{bfkl}).
By introducing an infra-red cutoff one can restrict the calculation to
contributions from the region where perturbation theory should safely apply.
We may also want to introduce an upper cutoff on $k^{2
\prime}_{\perp}$ for reasons of energy conservation \cite{cl,fh}.
 This is a necessary ingredient if one is to satisfy energy
conservation but certainly not sufficient when one sums over an
infinite number of gluons. We will return to a discussion of this question
later.

For a discussion and a numerical investigation of the effects of
introducing cutoffs on the BFKL equation see \cite{akms,fh,hr}. These papers
conclude that eq.(\ref{bfkl}) predicts a contribution to the gluon
distribution which rises like $x^{-K_0}$ up to powers of $\ln x$, (where
${\displaystyle K_0 = 12\,\alpha_s \ln2/\pi } $ ), i.e. that the small
$x$ slope is relatively stable to infra-red (and ultra-violet) cutoffs.
However, they also conclude that the
normalisation of this contribution is uncertain since it depends on the
details of the treatment of the infra-red region. Consequently, an exact
prediction for the contribution of the BFKL component
to $F_2(x,Q^2)$ is not possible.

Here we consider whether one can provide an analytic solution to the BFKL
equation with momentum cutoffs.
This has an advantage over numerical solutions in that it does not require
specification of the input distribution. Moreover the effects of the cutoffs
will be manifest. Our work is closely related to that of Collins and Landshoff
\cite{cl} but provides a different form for the solution which is more easily
generalised to deal with the exact kernel of the BFKL equation. It also
provides a formalism that allows us to investigate some of the implications
of energy conservation not dealt with by the momentum cutoff.

In a physical gauge two types of graph lead to equal powers of
$\alpha_s$ and $\ln(1/x)$: ladder graphs involving gluons with strong
ordering in $x$ up the ladder (the $f(x,k^{2\prime}_{\perp})$ terms
in eq.(\ref{bfkl})) and a subclass of the possible virtual
corrections to these graphs which reggeize the t-channel gluons
($f(x,k^2_{\perp})$ terms). Following the work of
Collins and Landshoff \cite{cl},  we restrict ourselves to introducing
cutoffs on  the real graphs only.
This makes the problem more tractable and ensures a gauge invariant result.

The equation is best solved by taking Mellin transforms with respect to
$x$ and $k^2_{\perp}$, solving the equation in double Mellin transform space
(DMT-space),  then inverting the solution back to
($x,k_{\perp}^2$)-space. The kernel of the BFKL equation has poles
in the left and right $\omega$-plane, $\omega$ being the transform
variable conjugate to $k_{\perp}^2$.
We present a method for deriving the DMT gluon distribution
in the cases where infra-red and/or ultra-violet cutoffs are imposed
upon the real terms (the $f(x,k^{2\prime}_{\perp})$) in eq.(\ref{bfkl}).
Our method is valid for any input distribution and can include any
number of the poles associated with the kernel. To illustrate the method we
give two examples, which we compare directly with the results
of ref.\cite{cl}.
In the example where both infra-red and ultra-violet cutoffs are
present, we do not agree with the results of ref.\cite{cl} (we differ by a
relative minus sign between terms). We discuss the effects upon the
small-$x$ behaviour of the gluon distribution function due to the
introduction of such cutoffs.

\section{The Pole Projection Method}

Mellin transforms with respect to $x$ and $k^2_{\perp}$
are defined by
\begin{eqnarray}
  f(x,k^2) &=&
             \int^{c+i\infty}_{c-i\infty} \frac{d\omega}{2\pi i}
             (k^2)^{\omega+\frac{1}{2}} \tilde{f}(x,\omega) \label{eq mt} \\
  \tilde{f}(x,\omega) & = &
               \int_{0}^{\infty} d{k^2}(k^2)^{-\frac{3}{2}-\omega}
               f(x,k^2) \label{eq imt}                        \\
  \tilde{f}(x,\omega)  & = & \int^{c+i\infty}_{c-i\infty}
		\frac{dN}{2\pi i}x^N {\cal F}(N,\omega) \label{eq mom}   \\
  {\cal F}(N,\omega) & = & \int^{1}_{0}dx x^{-1-N}
               \tilde{f}(x,\omega).
\end{eqnarray}

The contours in eqs.(\ref{eq mt}, \ref{eq mom}) lie parallel to the
imaginary axis and their position is chosen such that the respective
inverses exist. The addition of $1/2$ to the power of $k^2$
included in eq.(\ref{eq mt}) is merely for convenience.

In DMT-space the solution to the BFKL equation without momentum
cutoffs is given by a simple geometric sum:
\begin{equation}
{\cal F}(N,\omega)
= \sum_{n=0}^{\infty} \left( \frac{ -K(\omega) }{ N } \right)^n
{\cal F}_0(N,\omega) = \frac{1}{1+N^{-1}K(\omega)} {\cal F}_0(N,\omega)
\label{sum}
\end{equation}
where  $K(\omega)$ are the eigenvalues of the kernel of eq.(\ref{bfkl}),
\begin{equation}
K(\omega) = \frac{3\alpha_s}{\pi} [- 2\gamma - \psi(\frac{1}{2} + \omega) -
\psi(\frac{1}{2} - \omega) ],
\label{ksi}
\end{equation}
corresponding to the eigenfunctions $(k_{\perp}^2)^{1/2+\omega}$. The
quantities  $\psi$ and $\gamma$ are the log derivative of the Euler
gamma function and
the Euler gamma constant, respectively.

The eigenvalues given by eq.(\ref{ksi}) have an infinite set of simple poles
at $\omega$ = \{ $ \pm 1/2, \pm 3/2, \cdots $~\}
as a result of the poles in the $\psi$ functions. The contour  in
the transform definition eq.(\ref{eq mt}) is chosen to lie midway between
these poles and parallel to the imaginary axis.
We can expand $K(\omega)$ in terms of these poles

\begin{equation}
K(\omega) = \sum_{i}^{\infty} {\frac{a_i}{\omega-\omega_i}} +
            \sum_{j}^{\infty} {\frac{b_j}{\omega+\omega_j}} +
	    h(\omega)
\label{kpole}
\end{equation}
where $h(\omega)$ is an analytic function. For an analytic solution we will
approximate the infinite series by including only a
finite number of these poles (bearing in mind that the integral defined in
eq.(\ref{eq mt}) will, on closing the contour, be dominated by the nearest
singularities).

We turn now to the issue of imposing momentum cutoffs on the BFKL equation.
The important point to note is that the effect of introducing an infra-red
cutoff on the real graphs in eq.(\ref{bfkl})
is to remove the right-half plane poles of ${\cal F}(N,\omega)$ leaving only
those in the left-half $\omega$-plane.
To see this we use the following integral representation of the
$\Theta$-function\footnote{The contour lies to the right of
the pole at $\mu =0$ and is parallel to the imaginary axis.
 If $ k^2 > Q^2_0 $, then the contour is closed to the left, we pick up the
residue of the pole at $\mu=0$ and the result is 1. If $ k^2 < Q^2_0 $,
then closure is to the right, where there are no poles, and the
result is 0.} \cite{cl}
\begin{equation}
\Theta(
k^2 - Q^2_0
) = \int_c \frac{d \mu }{2 \pi i}
 \frac{ (k^2 / Q_0^2)^{\mu} }{ \mu }.
\label{theta}
\end{equation}

Let us define some function, $f_c (x,k^2_{\perp}) $,
which is non-zero only for $k^2_{\perp} > Q_0^2$,
(hereafter subscript `$c$' denotes a cutoff quantity)
in terms of an unrestricted function, $ g(x,k^2_{\perp}) $, by
\begin{equation}
f_c (x,k^2_{\perp}) \equiv
\Theta(k^2 - Q^2_0 ) g(x,k^2_{\perp}).
\label{defF}
\end{equation}
By taking double Mellin transforms of both sides we obtain the equation
\be
{\cal F}_c (N,\omega) =
\int_{c_{\nu}} {\frac{d\nu}{2 \pi i}
\frac{
Q_0^{ 2(\nu - \omega) } {\cal G}
 (N,\nu) }{\omega - \nu}}
\label{cc1}
\ee
where we have used eq.(\ref{theta}) and changed variables from $\mu$ to
$\nu= \omega -  \mu$. For the cases of interest
the asymptotic behaviour of ${\cal G}(N,\nu)$ is such that we can
close the contour via an infinite semicircle in either the left-half and/or
the right-half $\nu$-plane with no contribution from the contour at infinity.
If we take  ${\cal G}(N,\nu)$ to have only simple poles in the $\nu$-plane
then it follows that ${\cal F}_c (N,\omega)$ has poles only in the left-half
plane (the right-half plane poles being projected out by the theta function
which imposes the infra-red cutoff).  Conversely for an ultraviolet cutoff
only the right-half plane poles survive.

It will prove to be useful  in solving the BFKL integral equation to
 find an integral representation for ${\cal F}_c (N,\omega) $. Since
$f_c ( x,k^2_{\perp} ) = 0 $ for all
$ k^2 < Q^2_0 $, we may write
\begin{equation}
f_c (x,k^2_{\perp}) =
\Theta(k^2 - Q^2_0 )
f_c (x,k^2_{\perp}).
\end{equation}
Thus we have
\begin{equation}
{\cal F}_c (N,\omega) =
\int_{c_{\nu}} {\frac{d\nu}{2 \pi i}
\frac{ Q_0^{ 2(\nu - \omega) } ({\cal F}_c
 (N,\nu) - {\cal S}(N,\nu) ) }{\omega - \nu}}
\label{amb}
\end{equation}
and we have made explicit the fact that we may add a function
${\cal S}(N,\nu)$ to the integrand of the right-hand side which has
poles in
the right-half $\nu$-plane (provided that $S(N,\nu)$ has the asymptotic
behaviour which allows the contour to be closed in the left-half
$\nu$-plane with no contribution from the semicircle at infinity).

In the presence of an infra-red cutoff on the real emission terms, the BFKL
equation takes the form \cite{cl}:
\begin{eqnarray}
	{\cal F}_c(N,\omega) & = & {\cal F}_{0,c}(N,\omega)  \nonumber \\
	& + &
	{\displaystyle \int
        { \frac {d\nu}{2 \pi i} }
        { \frac{ Q_0^{2 (\nu-\omega) } } {\omega - \nu } }
         \left(  \frac{-K(\nu)}{N}  \right)
        {\cal F}_c (N,\nu)   }
	\label{eq cl15}
\end{eqnarray}
where ${\cal F}_{0,c}$ is the initial distribution with a cutoff.
We can now solve this integral equation by subtracting it
 from eq.(\ref{amb}) to give
\begin{equation}
	{\displaystyle \int
        { \frac {d\nu}{2 \pi i} }
        { \frac{ Q_0^{2 (\nu-\omega) } } {\omega - \nu } }
        ( {\cal F}_c(N,\nu) - S(N,\nu)  + (\frac{K(\nu)}{N}){\cal F}_c
	(N,\nu) - {\cal F}_{0,c}(N, \nu) ) }
        = 0.
\end{equation}

Since we have taken care to add ${\cal S}$ to describe the ambiguities
 we may equate the {\it integrand}  in
the above equation to zero. Thus
\begin{equation}
(1+N^{-1}K(\omega))
 {\cal F}_c(N,\omega)
=
{\cal F}_{0,c}(N,\omega) +
{\cal S}(N,\omega) ,
\label{proj}
\end{equation}
i.e.
\begin{equation}
 {\cal F}_c(N,\omega)
= {\displaystyle \frac{1}{(1+N^{-1}K(\omega))}
({\cal F}_{0,c}(N,\omega) +
{\cal S}(N,\omega) )} .
\label{cut}
\end{equation}
Comparing to the non-cutoff solution eq.(\ref{sum}) we see
that introducing the infra-red cutoff  has two effects.
Firstly, it converts the transform of the input solution to its cutoff
equivalent (which induces a factor $Q_0^{-2\omega}$ in ${\cal F}_{0,c}$),
and secondly it introduces the function ${\cal S}$ whose  right-half
plane poles cancel those present in the original solution ${\cal F}$
 leaving ${\cal F}_c(N,\omega)$ with only left-half plane poles
(this happens for each term in the geometric series). Note that as
$N^{-1}K(\omega)$ has only simple poles then ${\cal S}(N,\omega)$ will also
have only simple poles, as required. Moreover the asymptotic behaviour of
${\cal S}(N,\omega)$ must include the factor $Q_0^{-2\omega}$ to ensure the
correct behaviour of the integrand at infinity.

The right-half plane poles of ${\cal S}$
must balance those of $N^{-1}K(\omega){\cal F}_c$ on the left-hand side of
eq.(\ref{proj}).  If we keep $n$ right-half plane
poles in eq.(\ref{kpole}) then we will have $n$ conditions which will
determine ${\cal S}$. Thus we may write
\begin{equation}
{\cal S}(N, \omega) =
\sum_{i}^{n}{\frac{a_i^{\prime}}{\omega - \omega_i}}
\label{spoles}
\end{equation}
where $\omega_i = 1/2, 3/2, \cdots, (2n-1)/2$ and the
coefficients $a_{i}^{\prime}$ are determined by the cancellation
of the right-half plane poles in the denominator of eq.(\ref{sum}).
If $(1+N^{-1}K(\omega))$ has zeros at $\omega = \omega_s$ then
the $n$ conditions are
\begin{equation}
{\cal F}_{0}(N,\omega_s) +
{\cal S}(N,\omega_s) = 0
\label{conds}
\end{equation}
for the $n$ values of $s$. This procedure generates the solution to
eq.(\ref{eq cl15}).

Let us turn now to the case that there is both a lower and an
upper cutoff on $k_{\perp}^{2\prime}$. In this case the BFKL equation becomes
\begin{eqnarray}
{\cal F}_c(N,\omega) & = & {\cal F}_{0,c}(N,\omega) \nonumber \\
&+&
{\displaystyle \int
{ \frac {d\nu}{2 \pi i} }
{ \frac{ Q_0^{2 (\nu-\omega)}- Q_1^{2 (\nu-\omega) } } {\omega - \nu }}
\left( { \frac{-K(\nu)}{N} } \right)
{\cal F}_c (N,\nu)   }
\label{eq cl15p}
\end{eqnarray}
where $Q_1^2$ is the upper cutoff.The solution to this equation
proceeds in an analogous manner to that presented above.
We introduce the integral representation
\begin{equation}
{\cal F}_c (N,\omega) =
\int_{c_{\nu}} {\frac{d\nu}{2 \pi i}
\frac{
(Q_0^{ 2(\nu - \omega) }-Q_1^{ 2(\nu - \omega) }) ({\cal F}_c
 (N,\nu) - {\cal S}(N,\nu) ) }{\omega - \nu}}.
\label{amb2}
\end{equation}
As before, the function $S(N,\nu)$ can be any function that gives
zero after the integration of eq.(\ref{amb2}).
Since the first term (in parentheses) of the integrand is zero at
$\omega = \nu$ it follows that the function ${\cal S}(N,\nu)$ can have
left-half (right-half) poles providing that the asymptotic behaviour
of their associated functions allow the closure
of the integration contour in the right-half (left-half) $\nu$-plane
with zero contribution from the semicircle at infinity.
In order to determine ${\cal S}(N,\nu)$ we recall that a
function with a lower (upper) momentum cutoff has no poles in the right-half
(left-half) $\nu$-plane. Consequently ${\cal F}_c$ with both lower and
upper momentum cutoff must have no poles at all in the  $\nu$-plane.
Thus the function ${\cal S}(N,\nu)$ will be determined by the
condition that it remove all the poles of ${\cal F}_c$.
This procedure leads to the solution to eq.(\ref{eq cl15p}),
 again in the form of eq(\ref{cut}).

In general,  if we keep $n$ right-half and $m$ left-half plane
poles in $K(\omega)$ then we have $n+m$ conditions to
determine the $n+m$ coefficients in ${\cal S}$. With
\begin{equation}
{\cal S}(N, \omega)
=
\sum_{i}^{n}{\frac{a_i^{\prime}}{\omega - \omega_i}}  +
\sum_{j}^{m}{\frac{b_j^{\prime}}{\omega + \omega_j}}
\end{equation}
the conditions are the $n$ of eq.(\ref{conds})
together with $m$ new conditions given by
\begin{equation}
{\cal F}_{0}(N,-\omega_t) +
{\cal S}(N,-\omega_t) = 0.
\end{equation}
This completes the solution of the BFKL equation
for the case of an upper and a lower momentum cutoff.

\section{Examples}
We now illustrate the method by considering two examples:
\begin{enumerate}
\item {Infra-red cutoff including only the nearest poles in $K(\omega)$};
\item {Infra-red and ultra-violet cutoffs with only the nearest poles
       kept in $K(\omega)$}.
\end{enumerate}

\noindent We choose these two cases since they have been
studied already by Collins \& Landshoff (see ref.\cite{cl}) and so we may
compare the solutions derived by the two methods.
Thus we take
\begin{equation}
K(\omega) = {\displaystyle \frac{3 \alpha_s}{\pi}\frac{4 \ln
2}{(1-4\omega^2)} = \frac{K_0}{(1-4\omega^2)} }.
\label{kw}
\end{equation}

\subsubsection*{Example 1}

Introducing an infra-red cutoff on the BFKL equation removes the
possibility of right-half plane poles in ${\cal F}_c$ (see eq.(\ref{proj})).
Since we know ${\cal F}_c$ and ${\cal F}_{0,c}$ have no poles in the
right-half plane, the poles in  ${\cal S}$ must cancel those present in
the $N^{-1}K(\omega){\cal F}_c$ term, i.e.\ $ {\cal S} $ must have the same
right-half plane pole as $K(\omega)$, so
\[
{\cal S} = \frac{a_1}{ \omega-\frac{1}{2} }.
\]
The only zero of $ (1+N^{-1}K(\omega)) $ which lies in the right-half plane,
for the choice eq.(\ref{kw}), is at $\omega=\omega_N$ where
\[
 \omega_N = \frac{1}{2} \sqrt{1 + \frac{K_0}{N}}.
\]
\noindent For ${\cal F}_c$  not to have right-half plane poles
 it follows that
\begin{equation}
 {\cal F}_{0,c} (N,\omega_N) +  {\cal S}(N,\omega_N) = 0 .
\label{cond1}
\end{equation}
The function ${\cal S}$ required for this cancellation
is uniquely determined:
\[
 {\cal S} (N,\omega) = {\displaystyle
	\frac{ \frac{1}{2}-\omega_N }{\omega-\frac{1}{2}}
	{\cal F}_{0,c}(N,\omega_N) Q_0^{-2(\omega - \omega_N) } }.
\]

Note the inclusion of  the factor
 $ Q_0^{-2(\omega - \omega_N)} $ in ${\cal S}$ (which is unity when
$\omega=\omega_N$ for consistency with eq.(\ref{cond1})).
This gives ${\cal S}$
the correct  asymptotic behaviour (i.e. giving zero contribution to the
 contour integral from the semicircle at infinity).
Hence, we have for our cutoff solution
\begin{eqnarray}
  {\cal F }_c (N,\omega) & = & {\frac {1}{1 + N^{-1}K(\omega)}} \times
\nonumber \\
 &  & \left( {\displaystyle {\cal F}_{0,c}(N,\omega) +
 {\frac{ {\cal F}_{0,c}(N,\omega_N) (\frac{1}{2} - \omega_N)
           Q_0^{-2(\omega -  \omega_N)} }
                              {  \omega-\frac{1}{2}  } } }
                     \right),
\label{irpp}
\end{eqnarray}
which agrees with that derived in \cite{cl}, i.e.
\begin{eqnarray}
{\cal F}_c(N,\omega) & = & { \displaystyle\frac{1} {1+N^{-1}K(\omega_0) }}
\times \nonumber \\
& &
\left( {\cal F}_{0,c}(N,\omega) +
           { \displaystyle
               \frac{ {\cal F}_{0,c}(N,-\frac{1}{2})
		(\frac{1}{2} - \omega_N) }
                     { \omega + \omega_N }     }
             Q_0^{-1-2\omega}
\right) .
\label{ircl}
\end{eqnarray}

\subsubsection*{Example 2}

In this example we introduce an ultra-violet, $Q^2_1$, and an infra-red
cutoff, $Q^2_0$. Again, we keep only the nearest poles of the
eigenvalues so that $K(\omega)$ takes the form of eq.(\ref{kw}).
In this case,  the ultra-violet
cutoff projects out the remaining left-half plane poles of ${\cal
F}_{c}$.
We therefore insist that the function we add, $ {\cal S} $, to the original
solution cancels off poles in both the left-half and right-half of the
$\omega$-plane.  Since ${\cal S}$ must cancel the two poles of
$K(\omega)$ it must be of the form
\[
{\cal S}(N, \omega) = {\displaystyle \frac{a_1}{\omega - \frac{1}{2}} +
\frac{b_1}{\omega + \frac{1}{2} }} .
\label{sdc}
\]
The coefficients are determined by the fact that the poles in ${\cal F}_c$
associated with with the zeros of $(1+N^{-1}K(\omega))$ are
explicitly cancelled. We have two conditions
\begin{equation}
{\cal F}_{0,c}(N, \pm \omega_N)
 + {\cal S}(N, \pm \omega_N)
= 0 .
\end{equation}
${\cal S}(N,\omega)$  has terms with left-half (right-half) plane
poles which have the appropriate asymptotic behaviour to ensure
convergent closure of the contour in the right-half (left-half) $\nu$-plane.
Thus, after including the appropriate factors
$Q_0^{-2\omega}$ or $Q_1^{-2\omega}$, we arrive at the following
form for the solution in the double-cutoff case
\begin{eqnarray}
{\cal F}_{c}(N,\omega) & = &
{ \frac{1}{1 + N^{-1}K(\omega)} } \times \nonumber \\
 &  & \left[ {\cal F}_{0,c}(N,\omega)  +
{\frac{1}{\Delta(\omega_N)}} \left\{ \Delta(\omega_N){\cal S}(N,\omega)
 \right\} \right],
\label{eq dcpp}
\end{eqnarray}
where
\begin{eqnarray}
\Delta(\omega_N){\cal S}(N,\omega)
& = &
{\displaystyle
\frac{ Q_0^{-2\omega} }{ ( \omega - \frac{1}{2} ) } } \nonumber \\
 & \times  &
\left[
(\frac{1}{2} + \omega_N)
Q_1^{2\omega_N}{\cal F}_{0,c}(N,\omega_N) -
(\frac{1}{2} - \omega_N)
Q_1^{-2\omega_N} {\cal F}_{0,c}(N,-\omega_N) \right] \nonumber  \\
 & + &
{\displaystyle
\frac{ Q_1^{-2\omega} }{ ( \omega + \frac{1}{2} ) } } \nonumber \\
 & \times &
\left[
(\frac{1}{2} - \omega_N)
Q_0^{2\omega_N}{\cal F}_{0,c}(N,\omega_N) -
(\frac{1}{2} +  \omega_N)
Q_0^{-2\omega_N}{\cal F}_{0,c}(N,-\omega_N) \right] . \nonumber \\
& &
\label{deltas}
\end {eqnarray}
This may be rewritten
\begin{eqnarray}
{\cal F}_c (N,\omega) & = &
{ \frac{1}{1 + N^{-1}K(\omega_0)} } \times \nonumber \\
 &  & \left( {\cal F}_{0,c}(N,\omega)
-  G(\omega, \omega_N)
-  G(\omega, -\omega_N) \right),
\label{dccltrue}
\end {eqnarray}
with
\begin{eqnarray*}
 G(\omega,\omega_N) & = & {\displaystyle \frac
                          {(Q_0^{2(\omega_N - \omega)}
			 - Q_1^{2(\omega_N - \omega)} ) C(\omega_N)  }
			   {\omega - \omega_N}	}   \\
\Delta (\omega_N)C(\omega_N) & = & {\displaystyle \frac
				 { (\frac{1}{2} + \omega_N)Q_0^{1-2\omega_N}
			          {\cal F}_{0,c} (N,\frac{1}{2}) }
				 {R^{-2\omega_0} - R ^{-1}}  }  \\
& &                           - {\displaystyle \frac
				{ (\frac{1}{2} - \omega_N)Q_1^{-1-2\omega_N}
			        {\cal F}_{0,c}(N,-\frac{1}{2}) }
				{R^{2\omega_0} - R ^{-1}}  }, \\
\Delta(\omega_N) & = & {\displaystyle
 			{\frac
                        {\frac{1}{2} + \omega_N}{\frac{1}{2} - \omega_N} }
			R^{2 \omega_N}
                        - {\frac
        		{\frac{1}{2} - \omega_N}{\frac{1}{2} + \omega_N} }
			R^{-2 \omega_N} }
\end{eqnarray*}
where $R={\displaystyle \frac{Q_1}{Q_0}}$.
 In contrast, Collins and Landshoff found
\begin{eqnarray}
{\cal F}_{c} (N,\omega) & = &
 { \frac{1}{1 + N^{-1}K(\omega_0)} } \times \nonumber \\
 &  & \left( {\cal F}_{0,c}(N,\omega)
+ G(\omega, \omega_N)
+ G(\omega, -\omega_N) \right)
\label{eq dccl}
\end{eqnarray}

The two solutions (eq.(\ref{dccltrue}) and eq.(\ref{eq dccl})) differ
 only in a relative minus sign and
we have checked explicitly (by direct substitution into the integral equation)
that  eq.(\ref{dccltrue}) is the correct solution. Furthermore, by
taking the limit $Q_1 \to \infty $ in eq.(\ref{dccltrue}) we arrive at
eq.(\ref{ircl}) (this limit is not correctly reproduced from eq.(\ref{eq
dccl})).

\section {The solution in ($x,k^2_{\perp}$)-space}

Having produced a double cutoff solution ${\cal F}_{c}(N,\omega)$ that
is
free from poles one may inquire how it is that a non-zero gluon
distribution arises upon inverting back to ($x,k^2_{\perp}$)-space.
The key to understanding this lies in the inclusion of the $Q_0^{-2\omega}$
and $Q_1^{-2\omega}$ factors in ${\cal S}$.
Consider the two terms in eq.(\ref{deltas}) separately. Upon substituting
${\cal F}_{c}(N,\omega)$ in the inverse transform of eq.(\ref{eq imt}) the
first term forces the contour, $c_{\omega} $, to be closed to the left
(since ${\displaystyle k^2_{\perp}/Q_0^2 > 1 }$). On the other hand, the
second term must be closed to the right (since ${\displaystyle
k^2_{\perp}/Q_1^2 < 1 }$ ). In this way a non-zero function is
formed. It is easier in this case, however, to invert
eq.(\ref{dccltrue}) and arrive at
\begin{eqnarray}
{\cal F}_{c} (N, k^2_{\perp}) & = &  { \frac{-1}{1 +
N^{-1}K(\omega_0)} } \times  \\
& & {\displaystyle \left(\frac{A}{N + \epsilon}
(k^2_{\perp})^{\frac{1}{2} + \omega_0}
  + C(\omega_N)(k^2_{\perp})^{\frac{1}{2} + \omega_N}
  + C(-\omega_N)(k^2_{\perp})^{\frac{1}{2} - \omega_N} \right) }, \nonumber
\end {eqnarray}
for $Q_0 < k_{\perp} < Q_1 $ and 0 otherwise. Note this has an overall
minus sign when compared to eq.(22) of \cite{cl}.


We now discuss the structure of the solution in the N-plane and the
corresponding small-$x$ behaviour of the gluon density, for the
various cases.

In the non-cutoff case we have two poles at $\omega = \pm \omega_N$.
The contour, $c_{\omega}$, is pinched as these two poles move together
(corresponding to locating the nearest singularity in the $N$-plane)
and coincide at $\omega = 0 $. The $\omega$-plane contour in
eq.(\ref{eq mt}) is closed and this removes one of these poles to leave a
${\displaystyle 1/\sqrt{N+K_0}}$ singularity in the $N$-plane.
The $N$-plane contour is then closed and the discontinuity along this cut is
taken. This leads to the $x^{-K_0} \ln^{-1/2}(1/x)$ behaviour
(where $K_0 \simeq 1/2$ for $\alpha_s \simeq 0.2 $ (see eq.(\ref{kw})).

In the case of an infra-red cutoff we close the contour,
$c_{\omega}$, to the left and,  picking up the residue of the pole at
$\omega = -\omega_N$, we are left with the ($1/2 - \omega_N$)
terms explicit in eqs.(\ref{irpp}, \ref{ircl}) which then lead to a
$\sqrt{N+K_0}$ cut in the $N$-plane. The discontinuity across this cut leads
to a $x^{-K_0}\ln^{-3/2}(1/x)$ dependence. So introducing an infra-red
cutoff has made only a small difference to the $x$-dependence of the solution.

In the double cutoff case things are very different. The
singularities in the $N$-plane are at the zeros of $\Delta(\omega_N)$
see eqs.(\ref{eq dcpp} , \ref{dccltrue})
(with the exception of $\omega_N = 0$  which cancels).
This leads to a series of poles in the $N$-plane; the locations of which are
determined by solving the equation $\Delta(\omega_N) = 0$ and depend strongly
upon the value of $R$. The roots  of this equation
correspond to pure imaginary values of $\omega_N$. With $\bar{\omega}_N
= \Im m(\omega_N) $ the roots then satisfy the following equations:
\begin{eqnarray}
\bar{\omega}_N
& = &\frac{1}{2} \cot (\bar{\omega}_N \ln R) \hspace{0.6cm} {\rm{,and,}}
\label{cot} \\
\bar{\omega}_N
& =  &  - \frac{1}{2} \tan (\bar{\omega}_N \ln R) \label{tan}.
\end{eqnarray}
For large $R$, the solutions to these transcendental equations occur
near the zeros of the periodic functions and so we obtain approximate
solutions by assuming a linear approximation to these functions about
their zeroes. This leads to the solutions
\begin{equation}
\bar{\omega}_N \approx
\frac{n \pi/2}{2 + \ln R}
\label{wn}
\end{equation}
where $n$ is a positive integer.
The corresponding positions of the poles in the $N$-plane are given by
\begin{equation}
N = -K_0 {\displaystyle \left( \frac{1}{1 + 4 {\bar \omega}_N^2} \right)
\approx \frac{-K_0}{\left(
 1 + {\displaystyle \frac{n^2 \pi^2}{(2 + \ln R)^2} }
\right)} } .
\label{Npoles}
\end{equation}
The leading pole occurs for $n = 1$ in eq.(\ref{Npoles}) and for $R = 100$ is
$82\%$ of the asymptotic value (indeed it only falls to $65\%$ for $R$
as small as 10). We therefore disagree with the conclusion of
ref.\cite{cl} where the claim is that only for $R$ as big as $10^4$
does the position of the leading
singularity lie within $10\%$ of $-K_0$ and therefore that the necessary
cutoffs on the transverse momentum integrals have a `very significant
moderating effect' on the small-$x$ behaviour. The origin of the discrepancy
is clear and is merely a result of the approximations used in
ref.\cite{cl} when extracting the solutions to eq.(\ref{cot}).

The next-to-leading singularity ($n=2$) is determined by the solution to
eq.(\ref{tan}) and can lie close to the leading singularity for
moderate values of $R$, e.g. for $R=100$ it is $52\%$ of $-K_0$.
Of course as $R \to \infty$ all the poles merge together and form
the original square root branch cut.
We note that Collins \& Landshoff did not consider the solutions to
eq.(\ref{tan}) and consequently missed the location of this pole.

\section{Summary}
We have presented an analytic solution to the BFKL equation with infra-red
and/or ultra-violet cutoffs on the transverse momentum for real gluon
 emission. Our method is applicable to the general case where an
arbitrary number of poles in the kernel, $K(\omega)$, are included
and applies for any functional form for the input distribution.
Since our method leads to an expression for the solution in terms of
a geometric series (eqs.(\ref{sum},\ref{cut})) it may be modified to
allow for a summation over the emission of a finite
number of gluons \cite{fmr}. This should allow a more realistic implementation
of energy-momentum conservation.
\footnote{We thank Peter Landshoff for pointing
out the interest in such a restricted sum.}  We have verified the result of
Collins \& Landshoff in the case of an infra-red cutoff and correct their
result in the case of both infra-red and ultra-violet cutoffs. We find that
the imposition of both cutoffs reduces the small $x$ rise of the gluon density
but not by the amount claimed in ref.\cite{cl}. In particular, for an
infra-red cutoff of 1 GeV and an ultra-violet cutoff equal to the typical HERA
$\gamma p$ centre-of-mass energy, i.e. $\sim 100$ GeV, we expect a leading
power behaviour like $\sim x^{-0.4}$ (for $-K_0 = 0.5$). This behaviour is
consistent with current extractions of the gluon density from
HERA \cite{wolf,glue}.

\end{document}